\documentclass[preprint,aps]{revtex4}

\begin{document}

\title{Anomalous scaling of passive scalar in turbulence and in equilibrium}
\author{Gregory Falkovich and Alexander Fouxon}
\address{Physics of Complex Systems, Weizmann Institute of Science,
Rehovot 76100 Israel}


\begin{abstract}

We analyze multi-point correlation functions of a tracer in an
incompressible flow at scales far exceeding the scale $L$ at which
fluctuations are generated (quasi-equilibrium domain) and compare
them with the correlation functions at scales smaller than $L$
(turbulence domain). We demonstrate that the scale invariance can
be broken in the equilibrium domain and trace this breakdown to
the statistical integrals of motion (zero modes) as has been done
before for turbulence. Employing Kraichnan model of
short-correlated velocity we identify the new type of zero modes,
which break scale invariance and determine an anomalously slow
decay of correlations at large scales.
\end{abstract}

\maketitle

When the scale $L$ of an external source of fluctuations far
exceeds the scale at which fluctuations are dissipated, turbulent
cascade appears between those scales. One of the most interesting
fundamental aspects of turbulence is the existence of anomalies:
symmetries remain broken even when symmetry-breaking factors tend
to zero. For example, time-reversibility and scale invariance of
the statistics are not restored even when pumping scale goes to
infinity and dissipation scale to zero. The mechanism of
dissipative anomaly (responsible for irreversibility \cite{sree})
has been identified by Onsager as due to non-smoothness of the
velocity field in the inviscid limit \cite{onsa}. It is very much
similar to the axial anomaly in quantum field theory \cite{poly}.
Breakdown of scale invariance has been identified recently as
related to the statistical integrals of motion; in particular,
such integrals have been found as zero modes of the (multi-point)
operator of turbulent diffusion in the Kraichnan model of
short-correlated velocity \cite{SS1,GK,CFKL,SS,FGV}.

Here we consider passive scalar in a random incompressible flow
and ask whether the symmetries are restored at the scales far
exceeding the pumping scale. It is straightforward to see that
time reversibility holds (to put it simply, everything which is
pumped is dissipated at smaller scales and there is no flux
towards larger scales) \cite{FGV}. At the level of the second
moment or spectral density of the scalar, an equipartition takes
place at wave-numbers smaller than $L^{-1}$ like in other systems
with a direct cascade \cite{FGV,BFLS}. Here we find four-point and
higher moments at large scales and discover that the scale
invariance may be broken. We employ the Kraichnan model and
demonstrate that the breakdown is also due to statistical
integrals of motion (even though very much different from the zero
modes breaking scale invariance in turbulence). We thus find a
possible link between an anomalous scaling in equilibrium systems
(e.g. in critical phenomena) and in turbulence.

Consider the passive scalar $\theta({\bf r},t)$ carried by the
velocity ${\bf v}({\bf r},t)$ and pumped by
$\phi({\bf r},t)$:
\begin{equation}
\partial_t \theta+({\bf
v}\cdot\nabla)\theta=\phi\ .\label{ade}\end{equation}  The
characteristics of (\ref{ade}) are called Lagrangian trajectories
and defined by $\partial_t {\bf R}({\bf r}, t)= {\bf v}({\bf
R},t)$ and ${\bf R}({\bf r}, 0)={\bf r}$. Integrating (\ref{ade})
by characteristics we get $\theta({\bf r}, 0)=\theta ({\bf R}({\bf
r},-t),-t)+\int_{-t}^0\phi({\bf R}({\bf r},t_1), t_1)\,dt_1$.
Correlation functions, $F_n\!=\!\left\langle\overline{\theta_1
\ldots\theta_n}\right\rangle$, are obtained by averaging over 
velocity and pumping denoted by brackets and over-bar
respectively:
\begin{eqnarray}&&\!\!\!\!\!\!\!\! F_n\!=\!\bigl\langle \theta\left({\bf
R}_1(-t), -t\right)\dots
\theta\left({\bf R}_n(-t), -t\right) \bigr\rangle 
\!+\!\int\limits_{-t}^0\! dt_1\ldots\!\int\limits_{-t}^0 \!dt_n
\Bigl\langle \overline{\phi({\bf R}_1(t_1),t_1)\ldots \phi({\bf
R}_n(t_n),t_n})\Bigr\rangle \,.\label{forzato}
\end{eqnarray}
Here ${\bf R}_i(t)\equiv {\bf R}({\bf r}_i, t)$ and
$\theta_i=\theta({\bf r}_i,0)$. We assume that the pumping is
white Gaussian with a zero mean and the variance
$\overline{\phi({\bf r}_1, t_1)\phi({\bf r}_2, t_2)}=\chi({\bf
r}_{12})\delta(t_2-t_1)$, ${\bf r}_{ij}={\bf r}_i-{\bf r}_j$.

One can express the correlation functions via the multi-particle
propagators. For example, assuming zero conditions at the distant
past and space homogeneity, one gets
\begin{equation}
F_2({\bf r})=\int_{-\infty}^{0}dt\int P({\bf R},{\bf
r},t)\chi({\bf R})d{\bf R}\ ,\label{F2}
\end{equation}
where  $P({\bf R},{\bf r},t)$ is the probability density function
(pdf) of ${\bf R}_{12}(t)$ provided ${\bf R}_{12}(0)={\bf r}$,
${\bf R}_{ij}(t)={\bf R}_i(t)-{\bf R}_j(t)$.

We use the Kraichnan model \cite{Kra68,FGV} where velocity is
Gaussian with the zero mean and the variance
\begin{eqnarray} &&\langle v_\alpha({\bf
r}_1,t)v_\beta({\bf
r}_2,0)\rangle=\delta(t)\Bigl[K_0\delta_{\alpha\beta}-
K_{\alpha\beta}({\bf r}_{12})\Bigr]\,,\label{Kra}\\
&&K_{\alpha\beta}({\bf r})={Dr^{2-\gamma}\over d-1}
\left[(d+1-\gamma)\delta_{\alpha\beta}-(2-\gamma)
\frac{r_{\alpha}r_{\beta}}{r^2} \right]\ .\nonumber
\end{eqnarray}
Translation-invariant propagators satisfy closed differential
equations in the Kraichnan model \cite{Kra68,FGV}:
\begin{eqnarray}&&
\!\!\!\!\!\!\!\! (\partial_{t}+{\cal L}_n )P(\underline{\bf
R},\underline{\bf
 r},t)=0,\ \ P(\underline{\bf R},\underline{\bf
 r},0)=\delta(\underline{\bf R}-\underline{\bf r})\,,
 \label{l1}\end{eqnarray}
where $\underline{\bf R}=({\bf R}_1\ldots {\bf R}_n)$ and $2{\cal
L}_n\equiv \sum K_{\alpha\beta}({\bf
r}_{ij})\nabla_{i\alpha}\nabla_{j\beta}$. In particular, the pdf
of the two-particle separation
 [$P({\bf R},{\bf r},t)$ integrated over
angles] is expressed via the modified Bessel function
($\eta_0=d/\gamma-1$)
\begin{eqnarray}&&
\!\!\!\!\!\!\!\!
{\cal P}_0(R,r,{t})
\!=\!\frac{(Rr)^{-\gamma\eta_0/2}}{D\gamma
|t|}\exp\!\left[-\frac{R^{\gamma}\!+\!r^{\gamma}} {\gamma^2 |t| D
}\right]\!I_{\eta_0}\!\left(\frac{2(Rr)^{\gamma/2}} {\gamma^2 |t|
D }\right),\nonumber
\end{eqnarray}
giving the Richardson law of separation $R^\gamma\sim D |t|$
\cite{Kra68}.

We consider $\gamma>0$ (the case $\gamma=0$ has been described in
\cite{BFLL}), which corresponds to Holder exponent of the velocity
smaller than unity \cite{FGV}. In this case, the characteristics
are non-unique which is seen, for instance, from $P({\bf
R},0,t)\!\not=\!\delta({\bf R})$. Respective loss of information
leads to scalar dissipation, which balances pumping by $\phi$ and
provides for a statistical steady state of the scalar \cite{FGV}.
Spatially non-smooth velocities are produced by fluid turbulence
so that all scales considered in this paper are assumed to be less
than the outer scale of turbulence that is the correlation scale
of velocity. The equation (\ref{l1}) implies corresponding
equation on $F_{n}$:
\begin{equation}{\cal L}_{n}F_{n}=\sum_{i>j,k\neq i}
\chi({\bf r}_{ij})F_{n-2}(\{ {\bf r}_{k j} \})\
.\label{cf}\end{equation} The simplest steady state described by
(\ref{cf}) is the thermal equilibrium with Gaussian probability
density functional
$\propto\exp\left[ -\int \theta^2d{\bf r}/(2T)\right]$. Such state
has $F_2(r)=T\delta({\bf r})$ which requires also zero correlation
length $L$ of $\phi\,$:
\begin{eqnarray}&&
TK_{\alpha\beta}({\bf r})\nabla_{\alpha}\nabla_{\beta}\delta({\bf
r})=-\chi({\bf r})\,.\label{FDT}
\end{eqnarray}
That is if the limit $L\to 0$ is taken in such a way that $F_2 \to
T\delta({\bf r})$, the scalar statistics becomes Gaussian and
scale-invariant. The Gaussian anzatz, $F^G_{2n}\!=\!\sum F_2({\bf
r}_{ij})F^G_{2n-2}/{n}$ solves (\ref{cf}),
$$\sum
\nabla_i^\alpha\nabla_k^\beta\left[K_{\alpha\beta}({\bf
r}_{ik})-K_{\alpha\beta}({\bf r}_{jk}) \right]F_2({\bf
r}_{ij})F^G_{2n-2}
=0\,,$$
in two cases: i) $F_2({\bf r})=T\delta({\bf r})$ (in the
compressible case as well) and ii) $\gamma=2$ where
$K_{\alpha\beta}({\bf r})$ is ${\bf r}-$independent.

Of course, delta-function is an idealization so let us open a
Pandora box of anomalies by allowing $\phi$ to have a finite
correlation scale $L$. We assume that at $r\gg L$ the function
$\chi(r)$ decays faster than any power (say, exponentially). We
consider velocity field correlated at different points (i.e.
$\gamma<2$). Then, the scalar statistics is no longer Gaussian and
very different at the scales smaller and larger than $L$.

Let us first remind the properties of the turbulent state realized
at $r<L$. A nonzero flux $\chi(0)$ makes the pair correlation
function non-smooth at zero: $F(0)-F(r)\propto r^{\gamma}\chi(0)$.
The scaling properties are characterized by the structure
functions $S_{n}(r)=\langle [\theta({\bf
r})-\theta(0)]^{n}\rangle$ which are given by the zero modes of
the operator ${\cal L}_{n}$ \cite{GK,CFKL,CF,BGK}. Zero modes are
functions $Z\{{\bf R}_{ij}(t)\}$ of particles coordinates
conserved on average
 \cite{slow,FGV}.  The
structure functions are determined by the so-called irreducible
zero modes (that involve coordinates of all the points), their
exponents being $d\ln S_{n}/d\ln r=n\gamma/2-\Delta_n$ with the
anomalous exponents given by $\Delta_n=n(n-2)(2-\gamma)/2(d+2)$ in
the perturbative domain, $\Delta_n\ll n\gamma$. One can express
all the zero modes via the distance,
$R=\sqrt{R_{12}^2+\ldots+R_{1n}^2}$, and the angles,
$e_i=R_{1i}/R$, in $d(n-1)$-dimensional space:
$Z\{R_{ij}\}=R^\sigma f(\hat{\bf e})$. The conservation, ${\cal
L}Z=0$, in a Lagrangian language means that the growth of the
radial factor is compensated by the decay of the angular function.
The zero modes of the hermitian scale-invariant operator ${\cal
L}_{n}$ come in pairs satisfying the duality relation: for every
$Z_n=R^{\sigma_n} f_n(\hat{\bf e})$ there exists $Z_n'=$
$R^{\gamma-d(n-1)-\sigma_n} {\tilde f}_n(\hat{\bf e})$
\cite{slow}. There is the difference between the modes with
positive $\sigma$-s (that contribute at small scales) and with
negative $\sigma$-s (expected at large scales): ${\cal L}_{n}Z_n'$
produces delta-function at the origin or its derivatives (contact
terms) rather than zero. A physical reason is a nonzero
probability of initially distant particles to come to the same
point \cite{FGV}. For example, ${\cal L}_{2}R^{\gamma-d
}\propto\delta({\bf R})$ at $\gamma=2$.

After discovering that the zero modes $Z_n$ enter the solution of
 (\ref{cf}) at small scales, it is straightforward to ask what zero modes
 contribute at large scales where ${\hat L}_n F_n=0$.  Natural candidates to
contribute $F_{n}$ at $r>L$ are $Z_n'$ which decay with the
distances. And indeed this is true for $F_2$ for which the
following representation holds
\begin{eqnarray}&&
\!\!\!\!F_2={q_0r^{\gamma-d}\over D(d-\gamma)}
+\int_r^{\infty}\!\!\frac{\chi(y)y^{d-1}
(r^{\gamma-d}-y^{\gamma-d})dy}{D(d-\gamma)}\,.
\nonumber\end{eqnarray} Provided $q_0=\int \chi(r)\,d{\bf r}\neq
0$, zero mode $r^{\gamma-d}$ dual to the constant dominates $F_2$
at $r>L$. Such  duality between $r>L$ and $r<L$ does not work for
higher correlation functions. Considering, for instance, $F_4$ one
observes that the source appearing in (\ref{cf}) is a linear
combination of functions independent of one of three ${\bf
r}_{ij}$ on which $F_4$ depends. Since the action of ${\cal
L}_{n}$ on $Z_n'$ produces contact terms vanishing where any ${\bf
r}_{ij}$ is non-zero, then $Z_n'$ do not satisfy the correct
boundary conditions at $r\sim L$ and thus cannot contribute
$F_{n}$ at large scales. One can show that $Z_n'$ occur in the
solution only if irreducible terms are present in the pumping
correlation functions \cite{Fouxon}. That means that a Gaussian
pumping provides for $Z_n'$ only at small scales. Here we show
that the reducible terms in the pumping correlation functions
spawn a new type of zero modes at large scales, which correspond
to an axial (rather than spherical) symmetry in ${\bf R}$-space
and to the different duality relation,
$\sigma_n'=\gamma-nd/2-\sigma_k$.

Nonzero $q_0$ means infinite $\int F_2(r)\,d{\bf r}$ i.e an
overall heating.  We also call $q_0$ charge to invoke an analogy
with electrostatics (literal at $\gamma=2$ when ${\cal L}$ is a
Laplacian). When $q_0\neq 0$, $F_2(r)$ decays by a power law at
$r\gg L$ and high-order functions obey normal scaling though
statistics is non-Gaussian \cite{FGV,Fouxon}. We consider now a
pumping with $q_0=0$, like in (\ref{FDT}), when $F_2(r)$ decays
faster than any power. On a particle language, the distance ${\bf
R}(t)$ in (\ref{F2}) explores the sphere $R<L$ in a completely
isotropic manner with the result being proportional to $\int
\chi({\bf r})d{\bf r}$. Such a symmetry cancellation makes
pair-correlation small at large distances. We shall show now that
mutual correlation between different pairs of distances makes for
the power-law decays of multi-point correlation functions
(interpreted as dipole and quadrupole contributions).

The identity (\ref{forzato}) allows us to obtain two important
conclusions when all $r_{ij}\gg L$. Consider first $t\ll [\min
r_{ij}]^{\gamma}/D$, then the last term can be neglected since the
correlation function of $\phi$ is exponentially small. Averaging
over velocity at times earlier than $-t$ of which ${\bf R}_i(-t)$
are independent by zero correlation time of ${\bf v}$, one derives
that the correlation functions at large scales are statistical
invariants of Lagrangian dynamics:
\begin{eqnarray}&& F_n({\bf r}_1,\ldots
{\bf r}_n) \approx F_n\bigl({\bf R}_1(-t), \ldots
{\bf R}_n(-t)\bigr) \label{t2}\end{eqnarray} This is equivalent to
the above statement that $F_{n}$ is a zero mode of ${\cal L}_{n}$
at large distances. On the other hand, the steady state values of
the correlation functions can be obtained by taking in
(\ref{forzato}) the limit $t\to\infty$ when the first term
vanishes while the second one saturates. For Gaussian statistics
of $\phi$ this can only occur due to the rare events where pairs
of particles traced backwards in time come within the distance $L$
from each other. The distance between the pairs remains large with
overwhelming probability. Now one can choose intermediate times
such that (\ref{t2}) still holds but ${\bf R}_{i}(-t)$ are already
such that the  inter-pair distances is much smaller than the
distance between two pairs. This suggests that the main features
of, say, $F_4$ can be captured by considering the special geometry
of two distant pairs. In particular, this geometry will determine
the scaling of a coarse-grained field, see (\ref{final3}) below.
The overall scaling $\sigma_n$ of $F_{n}$ can also be inferred
from this particular case \cite{Fouxon}.

We consider  $F_4({\bf x},{\bf y},{\bf z})$, where ${\bf x}={\bf
r}_{12}$, ${\bf y}={\bf r}_{34}$ and ${\bf z}=({\bf r}_{13}+{\bf
r}_{24})/2$ at $x, y \ll z$. It is expressed via the pdf $P({\bf
X}, {\bf Y}, {\bf Z}, {\bf x}, {\bf y}, {\bf z}, t)$ of ${\bf
X}(t)={\bf R}_{12}(t)$, ${\bf Y}(t)={\bf R}_{34}(t)$ and ${\bf
Z}(t)=({\bf R}_{13}(t)+{\bf R}_{24}(t))/2$ conditioned at ${\bf
X}, {\bf Y}, {\bf Z}(t=0)={\bf x}, {\bf y}, {\bf z}$:
\begin{eqnarray}&&
\!\!\!\!\!\!\!\!\!\!\!\!\!\!\!F_4\approx\int_0^{\infty}\!\!\!\!
dt\, \int P({\bf X}, {\bf Y}, {\bf Z}, {\bf x}, {\bf y}, {\bf z},
t)H(X, Y)d{\bf X}d{\bf Y}d{\bf Z}\,. \label{o4}\end{eqnarray} Here
$H(X, Y)\equiv \chi(X)F_2(Y)+\chi(Y)F_2(X)$ decays where $max[X,
Y]> L$ because $q_0=0$. The integral is determined by
$t\simeq[\max\{x, y, L\}]^{\gamma}/D\ll z^{\gamma}/D$, that is the
distance $Z$ does not vary much. For such times, one can assume
$X\ll Z$, $Y\ll Z$ so that $X$ and $Y$ evolve approximately
independently. The lowest order approximation of the propagator is
via two-particles pdfs,
\begin{eqnarray}&&\!\!\!\!\!\!\!\!\!\!\!\! \!\!\!\int d{\bf Z}P({\bf X},
{\bf Y}, {\bf Z}, {\bf x}, {\bf y}, {\bf z}, t)\approx P({\bf X},
{\bf x}, t)P({\bf Y}, {\bf y}, t)\equiv P_0\,,
\nonumber\end{eqnarray} and it produces the reducible part of
$F_4$ that is $F_2(x)F_2(y)$, which is exponentially small if
$\max\{x, y\}>L$. The next-order corrections decay as powers of
$x/z$ and $y/z$ and thus dominate the correlation function if
$x>L$ or $y>L$. One can find that correction writing in (\ref{l1})
${\cal L}_4={\hat L}_0+{\hat L}_1$, ${\hat
L}_0=K_{\alpha\beta}({\bf
X})\nabla_{X_{\alpha}}\nabla_{X_{\beta}}+ K_{\alpha\beta}({\bf
Y})\nabla_{Y_{\alpha}}\nabla_{Y_{\beta}}$, see Appendix
\ref{sec:a31}. In the first order in ${\hat L}_1$ one derives
\begin{eqnarray}&&\!\!\!\!\!\!\!\!
P_1=\int_0^t dt'd{\bf x}'d{\bf y}'d{\bf z}'P({\bf X}, {\bf x}',
t')P({\bf Y}, {\bf y}', t')\delta({\bf Z}-{\bf
z}')
{\hat L}_1
P({\bf x'}, {\bf x}, t)P({\bf y'}, {\bf y}, t)\delta({\bf z}'-{\bf
z})\ . \nonumber
\end{eqnarray}
We now plug that into (\ref{o4}), use $\chi={\cal L}_2F_2$,
integrate by parts and obtain the first-order term in
$(x/z)^\gamma$, $(y/z)^\gamma$:
\begin{eqnarray}&&\!\!\!\!   F^1_4=\int_0^{\infty}
dt d{\bf x}'d{\bf y}'{\hat x}'_{\alpha}{\hat y}'_{\beta}F'_2(x')
F'_2(y')P({\bf x}', {\bf x}, t')
P({\bf y}',
{\bf y}, t') \left[x'_{\gamma}y'_{\delta} {\partial^2
K_{\alpha\beta}({\bf z})}/{\partial z_{\gamma}\partial
z_{\delta}}\right] \nonumber\end{eqnarray} The integration over
the directions of ${\bf x}'$ and ${\bf y}'$ gives ${\cal P}_2$
which is $P({\bf R},{\bf r},t)$ integrated over angles with the
second angular harmonics. It was found in \cite{CFKL} that ${\cal
P}_2$ has the same form as ${\cal P}_0$ with $\eta_0$ replaced by
$\eta_2=\gamma^{-1}\sqrt{(d-\gamma)^2+8d(d+1-\gamma)/(d-1)}$. Time
integration also can be done explicitly and we obtain (see
Appendix \ref{sec:a31} for the details of calculations)
\begin{eqnarray}&& \!\!\!\!\! F_4^1\!=\!\frac{(x y
)^{\gamma/4-d/2}}{2\pi\gamma D}{\hat x}_{\alpha}{\hat
x}_{\gamma}{\hat y}_{\beta}{\hat y}_{\delta}\frac{\partial^2
K_{\alpha\beta}({\bf z})}{\partial z_{\gamma}\partial
z_{\delta}}\int\limits_0^{\infty}\!\!dx'F'_2(x') 
\int\limits_{0}^{\infty}\!\!dy' F'_2(y')
(x'y')^{\gamma/4+d/2}Q_{\eta_2-1/2}(w)\,,\label{o15}
\end{eqnarray} where $Q_{\eta_2-1/2}$ is the Legendre
function of the second kind and $8w=(x y
x'y')^{-\gamma/2}[\left(x^{\gamma}+y^{\gamma}+x'^{\gamma}+y'^{\gamma}
\right)^2-4(xx')^{\gamma}-4(y y')^{\gamma}]$. The function $F_4^1$
approximates $F_4-F_2(x)F_2(y)$ in the whole region $z\gg L$, $x,
y\ll z$ irrespectively of $L$. At $x, y\gg L$ the correlation
function given by a zero mode of ${\cal L}_4$ reduces to the zero
mode of
${\hat L}_0$:
\begin{eqnarray}&&\!\!\!\!\!\!\!\!
F^1_4\propto\frac{(xy)^{\delta}q_2^2}
{(x^{\gamma}+y^{\gamma})^{(4\delta+2d-\gamma)/\gamma}}{\hat
x}_{\alpha}{\hat x}_{\gamma}{\hat y}_{\beta}{\hat
y}_{\delta}\frac{\partial^2 K_{\alpha\beta}({\bf z})}{\partial
z_{\gamma}\partial z_{\delta}}, \label{final0}\end{eqnarray} where
$2\delta=\gamma\eta_2+\gamma-d$ and $q_2\equiv \int
x^{\gamma+\delta}\chi(x)d{\bf x}$. We thus have shown that,
despite the fast decay of $F_2$ at large scales, $F_4$ decays as a
power-law there and found the scaling exponent:
$\sigma_4=2d+2\delta$. The dual zero mode of ${\hat L}_0$ with the
scaling exponent $\sigma_4'=\gamma-2d-\sigma_4$ (first found in
\cite{CFKL}) appears at $x, y\ll L$:
\begin{eqnarray}&&
Z^1\propto (xy)^{\delta}{\hat x}_{\alpha}{\hat x}_{\gamma}{\hat
y}_{\beta}{\hat y}_{\delta}\frac{\partial^2 K_{\alpha\beta}({\bf
z})}{\partial z_{\gamma}\partial z_{\delta}}\,.\nonumber
\end{eqnarray} One can easily verify for any vector
${\bf v}$ that $x^{\delta-2}[d({\bf x}\cdot{{\bf v}})^2-x^2v^2]$
is an anisotropic zero mode of the Lagrangian evolution operator
in the lowest order in the distance ${\bf x}$ between two
particles. The above derivation makes explicit the physical origin
of the anomalous scaling thus found: The main contribution into
$F_4$ at large scales comes from Lagrangian trajectories that were
in the past separated into two distant pairs. If the other pair
would not be there at all, the average pair prehistory would be
completely isotropic that is determined by $q_0=\int \chi(x)d{\bf
x}$ which vanishes. However, the presence of the other pair leads
to a preferred direction in space allowing each pair to exploit
the quadrupole term $q_2=\int \chi(x)x^{\gamma+\delta}d{\bf x}$.
Since the correlations between the pairs decay only as a power
law, the quadrupole contribution dominates the correlation
function. The possibility for more than two particles to explore
non-isotropic configurations (which show themselves via
non-isotropic zero mode) exists for higher correlation functions
as well.
To give an example of power-law correlations in high moments, let
us derive the explicit expression for $n-$point correlation
function in the limit of small $\xi=2-\gamma$. Besides reproducing
the correct scaling it shows how non-trivial the angular structure
of the correlation functions is at large scales. Since the
statistics of $\theta$ is Gaussian at $\xi=0$ we analyze the
cumulant 
$\Gamma_{2n}^1=\lim_{\xi\to 0}\Gamma_{2n}/\xi$ (see Appendix
\ref{sec:B}):
\begin{eqnarray}&&
\sum_k\nabla_k^2\Gamma_{2n}^1=F(n)
\sum_{\{i_j\}}\nabla_{i_1}^{\alpha}\nabla_{i_3}^{\beta}
\Bigl[J_{\alpha\beta}({\bf r}_{i_1i_3})+J_{\alpha\beta}({\bf
r}_{i_2i_4})\nonumber\\&& -J_{\alpha\beta}({\bf r}_{i_2i_3})
-J_{\alpha\beta}({\bf r}_{i_1i_4}) \Bigr]\prod_{k=1}^n
F_2\left({\bf r}_{i_{2k-1}i_{2k}}, \xi=0\right),\label{cum1}
\end{eqnarray} where the sum is over all permutations of indices,
$F(n)$ is an $n-$dependent
constant and
\begin{eqnarray}&&
J_{\alpha\beta}({\bf r})=D\ln r\delta_{\alpha\beta}-
{r_{\alpha}r_{\beta}}{r^{-2}}{D/(d-1)}\,.
\end{eqnarray}
The solution at $r_{ij}\gg L$ is as follows
\begin{eqnarray}&&
\!\!\!\!\!\!\!\!\Gamma_{2n}^1\approx
(-1)^{n+1}\frac{4\Gamma(nd/2+4)F(n)q_2^2q_1^{n-2}}{(\pi)^{nd/2}
D^{n-1}(d-1)(d+2)^3(2d)^n}
 \label{final1}\\&&\times
\sum_{\{i_j\}}\frac{f({\bf r}_{i_1}, {\bf r}_{i_2}, .., {\bf
r}_{i_{n}})}{c_{\{i_j\}}^{nd+4}},\ 
c_{\{i_j\}}^2={\sum_{k=1}^n\left({\bf r}_{i_{2k-1}}-{\bf
r}_{i_{2k}}\right)^2}\,, \nonumber \end{eqnarray} where $q_1=\int
x^\gamma\chi(x)\,d{\bf x}$ and a homogeneous function of zero
degree $f({\bf r}_1,..,{\bf r}_{2n})$ is  expressed via the
hypergeometric functions denoted $F_{ijk}=F(nd/2+4+i, j+1,
d/2+2+k, u)$ of the argument $u=-{2z^2}/{c^2}$ ($c^2\equiv
c_{\{1,2,..,2n\}}^2$) and the scalar products $a_1=\left({\bf
r}_{12}\cdot {\bf r}_{34}\right)/r_{12}r_{34}$, $a_2=({\bf
r}_{12}\cdot {\bf z})/r_{12}z$, $a_3=({\bf r}_{34}\cdot {\bf
z})/r_{34}z$:
\begin{eqnarray} &&f\!=\!\frac{r_{12}^2r_{34}^2}{2c^4}
\Biggl\{F_{000}\!+\!8a_2^2a_3^2\Biggl[\frac{u(nd\!+\!8)}{d+4}
F_{121}\!-\!F_{010} \!+\!F_{000}\Biggr]-a_1^2\left[2F_{010} +(d-2)
F_{000}\right] \nonumber \\&&
+2\left(a_2^2+a_3^2\right)\left(F_{010}-F_{000}\right)
-\!2a_1a_2a_3\Biggl[\frac{4u(nd\!+\!8)}{d+4}F_{121} 
\!+\!(d\!-\!4)(F_{010}\!-\!F_{000})\Biggr]\Biggr\}\,.\label{final2}
\end{eqnarray}
One can check that in the limit of large $z$ (\ref{final2})
reproduces (\ref{final0}) at $\xi\to 0$ (where $\delta\to 2$). Let
us stress that the cumulants $\Gamma_{n}$ are nonzero only when
the balance equation (\ref{FDT}) does not hold.
If to replace $\delta({\bf r})$ in (\ref{FDT}) by a function with
a finite $L$ then $q_0$ remains zero but dipole and quadrupole
moments, $q_1$ and $q_2$, are generally nonzero and
(\ref{final0},\ref{final1},\ref{final2}) appear.  The proof that
$\sigma_n=nd/2+2\delta$ for finite $\xi$ will be published
elsewhere \cite{Fouxon}.

To study how statistics changes with the scale one defines the
field coarse-grained over the scale $r$:
$\tilde\theta_\ell=\int_{r'<\ell}\theta({\bf r}')\,d{\bf
r}'/\ell^{d/2} $. A fast decay of the pair correlation function
together with power-law correlations of higher functions,
(\ref{final0},\ref{final1},\ref{final2}), mean that the statistics
of $\tilde\theta_\ell$ approaches Gaussian anomalously slow as
$\ell/L\to\infty$. Indeed, while the second moment
$\langle\tilde\theta_\ell^2\rangle$ tends to a constant
(exponentially) fast, the fourth cumulant may decay slower than
$\ell^{-d}$ (that one would have for fields with a finite
correlation radius):
\begin{eqnarray}&& \!\!\!\!\!\!\!\!\!\!\!\!\!\!\!\!
\langle\langle\tilde\theta_\ell^4\rangle\rangle \propto
\ell^{-d}\int c^{2d-1}dc\int F_4z^{d-1}dz\propto
(L/\ell)^{2\gamma}.\label{final3}\end{eqnarray} Here, the region
$c\simeq L\ll z\simeq \ell$ gives the main contribution  if
$2\gamma<d$ (see \cite{Fouxon} for details and also \cite{proc}).
Similarly one can derive
$\langle\langle\theta^2(0)\theta^2(r)\rangle\rangle\propto
r^{-2\gamma}$. Note that $F_4^1$ gives zero contribution and one
must account for $F_4^2\propto z^{-2\gamma}$.

Since $\sigma_n/n$ depends on $n$ then the statistics is not
scale-invariant at $r>L$ when $q_0=0$. If $q_0\not=0$ then the
zero modes found here provide for an anomalous scaling of
sub-leading corrections, more similar to what is generally
observed in critical phenomena. What we believe is of importance
here is that we trace this anomalous scaling to zero modes. That
shows that at least in passive scalar problem, the mechanism of an
anomalous scaling is common for turbulence and thermal equilibrium
and raises an intriguing possibility that in other problems in
quantum field theory and statistical physics one can relate
anomalous scaling to statistical integrals of motion.

Our work benefited from helpful remarks of K. Gaw\c{e}dzki, V.
Lebedev and A. Zamolodchikov. We are grateful to the participants
of the Eilat workshop (October 2004) for the discussion of the
results. After the workshop, the numerical simulations were
undertaken \cite{CS}, which seem to support (\ref{final3}) --- we
are grateful to A. Celani and A. Seminara for sharing their
results prior to publication. This research has been supported by
the Minerva grant 8464 and the EU Network "Fluid mechanical
stirring and mixing: the lagrangian approach".

\appendix\section{Calculation of $F_4$ at a special geometry}
\label{sec:a31}

The expression for the four-point correlation function $F_4({\bf
r}_1, {\bf r}_2, {\bf r}_3, {\bf r}_4)$ follows from
(\ref{forzato})
\begin{eqnarray}&&
F_4=\int_{-\infty}^0 dt_1dt_2 \Biggl\langle\Biggl[\chi[
R_{12}(t_1)]\chi[R_{34}(t_2)]+ \nonumber \\&& \chi[ R_{13}(t_1)]
\chi[R_{24}(t_2)]+\chi[
R_{14}(t_1)]\chi[R_{23}(t_2)]\Biggr]\Biggr\rangle.
\label{forzato1}\end{eqnarray} Here we consider calculation of
$F_4$ for the case where the absolute values of the inter-pair
distances ${\bf x}={\bf r}_1-{\bf r}_2$ and ${\bf y}={\bf
r}_3-{\bf r}_4$ are much smaller than the absolute value of the
distance ${\bf z}=[{\bf r}_1+{\bf r}_2-{\bf r}_3-{\bf r}_4]/2$
between the centers of mass of the pairs. We shall also assume
that $z\gg L$. It is easy to verify that the first summand
dominates (\ref{forzato1}) in this case. Indeed, the other two
terms are contributed by the events on which $z$ shrinks to the
distance of order $L$. These events have smaller probability than
similar events involving $x$ and $y$ justifying the above
statement. Finally, using zero-correlation time and
time-reversibility of Kraichnan model and employing $H(x,
y)=\chi(x)F_2(y)+\chi(y)F_2(x)$ introduced in the main text we
obtain
\begin{eqnarray}&&
F_4\approx \int_0^{\infty} dt \langle H\left(R_{12}(t),
R_{34}(t)\right)\rangle=\int_0^{\infty} dt H(X, Y) \nonumber
\\&& \times P({\bf X}, {\bf Y}, {\bf Z}, {\bf x}, {\bf y}, {\bf z}, t)
d{\bf X}d{\bf Y}d{\bf Z},
\label{04}\end{eqnarray} where we introduced the joint probability
density function $P({\bf X}, {\bf Y}, {\bf Z}, {\bf x}, {\bf y},
{\bf z}, t)$ of  ${\bf X}(t)={\bf R}_1(t)-{\bf R}_2(t)$, ${\bf
Y}(t)={\bf R}_3(t)-{\bf R}_4(t)$ and ${\bf Z}=[{\bf R}_1(t)+{\bf
R}_2(t)-{\bf R}_3(t)-{\bf R}_4(t)]/2$. In Kraichnan model this
function satisfies a closed evolution equation
\begin{eqnarray}&&
\partial_t P=[{\hat L}_0+{\hat L}_1]P,\ \
P(t=0)=\delta\left({\bf X}-{\bf x}\right)\delta\left({\bf Y}-{\bf
y}\right)\nonumber \\&& \delta\left({\bf Z}-{\bf z}\right),\ \
{\hat L}_0\!=\!K_{\alpha\beta}({\bf X})\frac{\partial^2 }{\partial
X_{\alpha}\partial X_{\beta}}+K_{\alpha\beta}({\bf
Y})\frac{\partial^2 }{\partial Y_{\alpha}\partial
Y_{\beta}},\nonumber \\&& {\hat L}_1=\frac{1}{4}
\Biggl[K_{\alpha\beta}\left({\bf Z}+\frac{{\bf X}-{\bf
Y}}{2}\right)+K_{\alpha\beta}\left({\bf Z}+\frac{{\bf X}+{\bf
Y}}{2}\right) \nonumber \\&& +K_{\alpha\beta}\left({\bf
Z}-\frac{{\bf X}+{\bf Y}}{2}\right)+ K_{\alpha\beta}\left({\bf
Z}+\frac{{\bf Y}-{\bf X}}{2}\right) \nonumber \\&&
-K_{\alpha\beta}({\bf Y})-K_{\alpha\beta}({\bf
X})\Biggr]\frac{\partial^2 }{\partial Z_{\alpha}\partial
Z_{\beta}} +\Biggl[K_{\alpha\beta}\left({\bf Z}+\frac{{\bf X}+{\bf
Y}}{2}\right) \nonumber \\&& +K_{\alpha\beta}\left({\bf
Z}-\frac{{\bf X}+{\bf Y}}{2}\right)-K_{\alpha\beta}\left({\bf
Z}+\frac{{\bf X}-{\bf Y}}{2}\right)\nonumber
\\&&-K_{\alpha\beta}\left({\bf
Z}+\frac{{\bf Y}-{\bf X}}{2}\right) \Biggr] \frac{\partial^2
}{\partial X_{\alpha}\partial Y_{\beta}}
+\frac{1}{2}\Biggl[K_{\alpha\beta}\left({\bf Z}+\frac{{\bf X}+{\bf
Y}}{2}\right) \nonumber
\\&& -K_{\alpha\beta}\left({\bf Z}-\frac{{\bf X}+{\bf
Y}}{2}\right)+K_{\alpha\beta}\left({\bf Z}+\frac{{\bf X}-{\bf
Y}}{2}\right) \nonumber \\&& -K_{\alpha\beta}\left({\bf
Z}+\frac{{\bf Y}-{\bf X}}{2}\right) \Biggr]\frac{\partial^2
}{\partial X_{\alpha}\partial Z_{\beta}}+\frac{1}{2}
\Biggl[K_{\alpha\beta}\left({\bf Z}+\frac{{\bf X}+{\bf
Y}}{2}\right)\nonumber
\\&& -K_{\alpha\beta}\left({\bf Z}-\frac{{\bf X}+{\bf Y}}{2}\right)
 +K_{\alpha\beta}\left({\bf Z}+\frac{{\bf Y}-{\bf
X}}{2}\right)\nonumber
\\&& -K_{\alpha\beta}\left({\bf Z}+\frac{{\bf X}-{\bf Y}}{2}\right)
\Biggr]\frac{\partial^2 }{\partial Y_{\alpha}\partial Z_{\beta}}.
\nonumber\end{eqnarray} Next we observe that the integral
(\ref{o4}) is determined by times of the order $[\max{x, y,
L}]^{\gamma}/D$ (we assume $q_0=0$ here so that $F_2$ is small
outside $L$). During these times which are much smaller
$z^{\gamma}/D$ the distance $Z$ does not vary much while $X$ and
$Y$ remain much smaller than $Z$. The main dynamics at these times
is just independent evolution of "fast" degrees of freedom ${\bf
X}$, ${\bf Y}$ governed by the operator ${\hat L}_0$. Considering
${\hat L}_1$ as a perturbation we construct $P$ as a series
\begin{eqnarray}&&
P=\sum_{n=0}^{\infty} P_n,\ \ \frac{\partial P_n}{\partial
t}={\hat L}_0P_n+{\hat L}_1P_{n-1},
\end{eqnarray}
where $P_0$ is given by $P_0({\bf X}, {\bf x}, t)P_0({\bf Y}, {\bf
y}, t)\delta({\bf Z}-{\bf z})$.  Here $P_0({\bf X}, {\bf x}, t)$
satisfies the initial condition $P_0(t=0)=\delta\left({\bf X}-{\bf
x}\right)$ and the evolution equation
\begin{eqnarray}&&
\frac{\partial P_0({\bf X}, {\bf x}, t)}{\partial
t}=K_{\alpha\beta}({\bf X})\frac{\partial^2 P_0}{\partial
X_{\alpha}\partial X_{\beta}}.\label{o3}\end{eqnarray} The
explicit solution for $P_n$ with $n>0$ reads
\begin{eqnarray}&&
P_n=\int_0^t dt'd{\bf x}'d{\bf y}'d{\bf z}'P_0({\bf X}, {\bf x}',
t-t')P_0({\bf Y}, {\bf y}', t-t') \nonumber \\&& \times
\delta({\bf Z}-{\bf z}')\left({\hat L}_1\right)_{x', y', z'}
P_{n-1}({\bf x}', {\bf y}', {\bf z}', t') \nonumber
\end{eqnarray}
Above the subscript of Hermitian operator ${\hat L}_1$ signifies
that it acts on variables ${\bf x}'$, ${\bf y}'$, ${\bf z}'$. The
asymptotic series for $P$ integrated according to (\ref{o4})
produces asymptotic series $F_4=\sum F^n$, where $F^{n+1}/F^n$
tends to zero as $z\to\infty$. To write this series most concisely
we note that
\begin{eqnarray}&&
\int d{\bf X}d{\bf Y}P_0({\bf X}, {\bf x}', t')P_0({\bf Y}, {\bf
y}', t')\Biggl[\chi(X)F_2(Y)+\chi(Y) \nonumber \\&& \times
F_2(X)\Biggr]=-\int d{\bf X}d{\bf Y}P_0({\bf X}, {\bf x}', t')
P_0({\bf Y}, {\bf y}', t')\Biggl[F_2(Y) \nonumber \\&& \times
K_{\alpha\beta}({\bf X})\frac{\partial^2 F_2(X)}{\partial
X_{\alpha}\partial X_{\beta}}+F_2(X)K_{\alpha\beta}({\bf
Y})\frac{\partial^2 F_2(Y)}{\partial Y_{\alpha}\partial
Y_{\beta}}\Biggr] \nonumber
\\&& =- \frac{\partial}{\partial t'}\int d{\bf X}d{\bf Y}P_0({\bf
X}, {\bf x}', t') P_0({\bf Y}, {\bf y}', t')F_2(X)F_2(Y),
\nonumber\end{eqnarray} where we used the equation satisfied by
$F_2$. Using the above we find the following expression for $F^n$
\begin{eqnarray}&&
F^n=\int_{0}^{\infty}dt \int P_n({\bf X}, {\bf Y}, {\bf Z}, {\bf
x}, {\bf y}, {\bf z}, t)H(X, Y)d{\bf X}d{\bf Y}d{\bf Z} \nonumber
\\&& =\int_0^{\infty} dt d{\bf x}'d{\bf y}'d{\bf z}'F_2(x')
F_2(y')\left({\hat L}_1\right)_{x', y', z'} P_{n-1}({\bf x}', {\bf
y}', {\bf z}', t)
\end{eqnarray}
Using the expression for $P_n$ we find the expression for $F^n$
with $n>2$ in terms of $F^1$:
\begin{eqnarray}&&
F^n=\int_0^{\infty} dt d{\bf x}'d{\bf y}'d{\bf z}'F^1{\hat
L}_1P_{n-2}({\bf x}', {\bf y}', {\bf z}', t).
\end{eqnarray} In particular, for $F^2$ we obtain
\begin{eqnarray}&&
F^2=\int_0^{\infty} dt d{\bf x}'d{\bf y}'P_0({\bf x}, {\bf x}',
t')P_0({\bf y}, {\bf y}', t'){\hat L}_1F^1({\bf x}', {\bf y}',
{\bf z}). \nonumber\end{eqnarray} We pass to study $F^1$ which can
be written as
\begin{eqnarray}&&
F^1=\int_0^{\infty} dt d{\bf x}'d{\bf y}'P_0({\bf x}, {\bf x}',
t')P_0({\bf y}, {\bf y}', t'){\hat x}'_{\alpha}{\hat
y}'_{\beta}F'_2(x') \nonumber \\&& \times
F'_2(y')\Biggl[K_{\alpha\beta}\left({\bf z}+\frac{{\bf x}'+{\bf
y}'}{2}\right)+K_{\alpha\beta}\left({\bf z}-\frac{{\bf x}'+{\bf
y}'}{2}\right)\nonumber
\\&&-K_{\alpha\beta}\left({\bf
z}+\frac{{\bf x}'-{\bf y}'}{2}\right)-K_{\alpha\beta}\left({\bf
z}+\frac{{\bf y}'-{\bf x}'}{2}\right) \Biggr].
\end{eqnarray}
Expanding the difference in brackets at large $z$ we obtain
\begin{eqnarray}&&  F^1=\int_0^{\infty} dt d{\bf x}'
d{\bf y}'P_0({\bf x}, {\bf x}',
t')P_0({\bf y}, {\bf y}', t'){\hat x}'_{\alpha}{\hat
y}'_{\beta}F'_2(x') \nonumber \\&& \times F'_2(y')
\left[x'_{\gamma}y'_{\delta} \frac{\partial^2 K_{\alpha\beta}({\bf
z})}{\partial z_{\gamma}\partial
z_{\delta}}+O\left(z^{-\gamma-2}\right)\right]
\nonumber\end{eqnarray} It is convenient to perform angular
integration by substituting the integrand by its value averaged
independently over the directions of ${\bf x}'$ and ${\bf y}'$. We
introduce functions averaged over the directions of ${\bf x}'$:
\begin{eqnarray}&&
\langle {\hat x}'_{\alpha}{\hat x}'_{\gamma}P_0({\bf x}, {\bf x}',
t)\rangle_{angle}=\delta_{\alpha\gamma} A+{\hat x}_{\alpha}{\hat
x}_{\gamma} B.
\end{eqnarray}
Taking the trace of the above equation and multiplying it with
${\hat x}_{\alpha}{\hat x}_{\gamma}$ leads to
\begin{eqnarray}&&
\langle P_0({\bf x}', {\bf x}, t)\rangle_{angle}=dA+B,\ \nonumber
\\&& \langle \left({\hat x}'\cdot {\hat x}\right)^2P_0({\bf x}', {\bf
x}, t)\rangle_{angle}=A+B.
\end{eqnarray}
Isotropy of velocity statistics implies that the LHS of the above
equations are independent of ${\hat x}$ so that $A=A(x, x', t)$,
$B=B(x, x', t)$. It follows from the above relations that
\begin{eqnarray}&&
(1-d)B=\langle \left[1-d\left({\hat x}'\cdot {\hat
x}\right)^2\right]P_0({\bf x}, {\bf x}', t)\rangle_{angle}.
\end{eqnarray}
Using incompressibility we can write the answer in terms of the
function $B$ solely
\begin{eqnarray}&&  F^1={\hat x}_{\alpha}{\hat x}_{\gamma}
{\hat y}_{\beta}{\hat
y}_{\delta}\frac{\partial^2 K_{\alpha\beta}({\bf z})}{\partial
z_{\gamma}\partial z_{\delta}}\int_0^{\infty} dt d{\bf x}'d{\bf
y}'B( x, x', t)  \nonumber \\&& B(y, y', t) x'y'F'_2(x') F'_2(y').
\end{eqnarray}
The function $B(x, x', t)$ can be found by expanding $P_0({\bf x},
{\bf x}', t)$ in Jacobi polynomials in ${\bf x}\cdot {\bf
x}'/(xx')$, see \cite{CFKL,Fouxon}. One finds that $B$ is the
coefficient near the Jacobi polynomial of the second-order and it
is given by \cite{CFKL}
\begin{eqnarray}&&
B(r, r', t)=\frac{(rr')^{-d/2+\gamma/2}}{S^{d-1}D(d-1)\gamma
t}\exp\left[-\frac{r^{\gamma}+r'^{\gamma}}{\gamma^2 (d-1)t
D}\right] \nonumber \\&& I_{\eta_2}\left(\frac{2
(rr')^{\gamma/2}}{\gamma^2 (d-1)t D}\right),\
\eta_2\!=\!\sqrt{\left(\!\frac{d}{\gamma}\!-\!1\right)\!^2\!+\!
\frac{8d(d\!+\!1\!-\!\gamma)}{\gamma^2\!
(d\!-\!1)} }. \nonumber\end{eqnarray} Up to a proportionality
factor the above coincides with $R_1$ from \cite{CFKL}. The
difference due to the misprint in \cite{CFKL} is discussed in more
detail in \cite{Fouxon}. The integral over time is given by
\cite{Brychkov}
\begin{eqnarray}&&
\int_0^{\infty}B( x', x, t)  B(y', y, t)dt=\frac{(x y
x'y')^{\gamma/4-d/2}}{(2\pi)(S^{d-1})^2(d-1) D} \nonumber \\&&
Q_{\eta_2-1/2}\left(\!
\frac{\left(x^{\gamma}\!+\!y^{\gamma}\!+\!x'^{\gamma}\!
+\!y'^{\gamma}\right)^2\!-\!4(xx')^{\gamma}\!-\!4(y
y')^{\gamma}}{8(x y x'y')^{\gamma/2}}\!\right).
\nonumber\end{eqnarray} The above leads to the result (\ref{o15})
from the main text. The analysis of the expression proceeds along
the following lines. We first consider $F^1$ where $x\gg L$ and
$y\gg L$. In this situation one can assume $x'\gg x$ and $y'\gg y$
and use
\begin{eqnarray}&&
Q_{\eta_2-1/2}\left(
\frac{\left(x^{\gamma}+y^{\gamma}+x'^{\gamma}+y'^{\gamma}\right)^2
-4(xx')^{\gamma}-4(y
y')^{\gamma}}{8(x y x'y')^{\gamma/2}} \right)\nonumber \\&&
\approx
\frac{\Gamma(\eta_2+1/2)\Gamma(1/2)}{2^{\eta_2+1/2}\Gamma(\eta_2+1)}
\left(
\frac{8(x y
x'y')^{\gamma/2}}{\left(x^{\gamma}+y^{\gamma}\right)^2}
\right)^{\eta_2+1/2},
\end{eqnarray}
where we used the large argument asymptotic form of the Legendre
function of the second kind. Substituting the above into the
expression for $F^1$ we find
\begin{eqnarray}&&
F^1=\frac{(xy)^{\delta}{\hat x}_{\alpha}{\hat x}_{\gamma}{\hat
y}_{\beta}{\hat
y}_{\delta}}{(x^{\gamma}+y^{\gamma})^{(4\delta+2d-\gamma)/\gamma}}
\frac{\partial^2
K_{\alpha\beta}({\bf z})}{\partial z_{\gamma}\partial
z_{\delta}}\frac{2^{2\eta_2}}{\pi\gamma d} \nonumber
\\&& \left(\int_{0}^{\infty}F'_2(x')x'^{\delta+d}dx'\right)^2
\frac{\Gamma(\eta_2+1/2)\Gamma(1/2)}{\Gamma(\eta_2+1)} .
\end{eqnarray}
In the opposite limit $x\ll L$ and $y\ll L$ one finds zero mode
dual to the above. Indeed, in this situation one can use the
inverse inequalities $x\ll x'$, $y\ll y'$ and
\begin{eqnarray}&&
Q_{\eta_2-1/2}\left(
\frac{\left(x^{\gamma}+y^{\gamma}+x'^{\gamma}+y'^{\gamma}\right)^2
-4(xx')^{\gamma}-4(y
y')^{\gamma}}{8(x y x'y')^{\gamma/2}} \right) \nonumber \\&&
\approx
\frac{\Gamma(\eta_2+1/2)\Gamma(1/2)}{2^{\eta_2+1/2}\Gamma(\eta_2+1)}\left(
\frac{8(x y
x'y')^{\gamma/2}}{\left(x'^{\gamma}+y'^{\gamma}\right)^2}
\right)^{\eta_2+1/2}.
\end{eqnarray}
One obtains
\begin{eqnarray}&&
F^1=\frac{2^{2\eta_2}\Gamma(\eta_2+1/2)\Gamma(1/2)}{\pi\gamma
d\Gamma(\eta_2+1)}(xy)^{\delta}{\hat x}_{\alpha}{\hat
x}_{\gamma}{\hat y}_{\beta}{\hat y}_{\delta}\frac{\partial^2
K_{\alpha\beta}({\bf z})}{\partial z_{\gamma}\partial z_{\delta}}
\nonumber \\&& \times
\int_0^{\infty}dx'\int_{0}^{\infty}dy'F'_2(x') F'_2(y')
\frac{(x'y')^{\delta}}{(x'^{\gamma}+y'^{\gamma})^{(4\delta+2d-\gamma)/
\gamma}}.\nonumber\end{eqnarray}
Note that now the dependencies on
$x'$ and $y'$ in the integral cannot be separated.

It is easy to see that $F^1$ vanishes if $x$ or $y$ or both are
zero. Indeed, if $x=0$ then by isotropy $P_0({\bf x}', 0,
t)=P_0(x', t)$.  The angular integration then produces zero by
identities like
\begin{eqnarray}&&
\int_{x=R} {\hat x}_{\alpha}K_{\alpha\beta}\left({\bf
z}+\frac{{\bf x}+{\bf y}}{2}\right) dS=0.
\end{eqnarray}
The same argument shows that $F^1$ averaged over the directions of
${\bf x}$ or ${\bf y}$ or both vanishes as well. As a result such
objects as $\langle\langle\theta^2(0)\theta^2({\bf
r})\rangle\rangle$ and $\langle\langle\theta_r^4\rangle\rangle$
are determined by $F^2$. These are calculated in \cite{Fouxon}.

\section{$n-$th order correlation functions in the limit of small
$\xi$}\label{sec:B}

We remind that passive scalar statistics is Gaussian at
$\xi=2-\gamma=0$. Thus it is convenient to discuss the limit of
small $\xi$ in terms of irreducible correlation functions that are
proportional to $\xi$ in this limit.

The analysis proceeds as follows. We first derive equations
satisfied by the irreducible correlation functions $\Gamma_{n}$ in
the limit of small $\xi$. We show that $\Gamma_{n}$ solves the
potential problem (electrostatics) in the space of $nd$ dimensions
with the source concentrated within a generalized cylinder. After
the integration over the directions parallel to the cylinder axis
we express $\Gamma_n$ as an integral of a kernel with localized
source. The large-scale asymptotic expansion of $\Gamma_{n}$ is
then derived exactly in the same way as the multi-pole expansion
in electrostatics with the coefficients given by two quadrupole
and $n/2-2$ dipole moments of the pumping correlation function. We
proceed to the calculation.

\subsection{Equations on $\Gamma_n$ in the limit of small $\xi$}

We stressed above that the statistics of $\theta$ is nearly
Gaussian at small $\xi$ so that $\Gamma_n$ are a small correction
to the reducible part of the correlation function which is
expressible via $F_2$. Thus the source appearing in the equation
on the leading order term in $\Gamma_n$ should be expressible in
terms of $F_2$. It is simplest to show this for
$\Gamma_4=F_4-F_4^G=F_4-F_2(r_{12})F_2(r_{34})-F_2(r_{13})F_2(r_{24})
-F_2(r_{14})F_2(r_{23})$. Using the equation on $F_4$ it is easy
to show that
\begin{eqnarray}&&
-\sum_{i>j}K_{\alpha\beta}({\bf r}_i-{\bf
r}_j)\nabla_{i\alpha}\nabla_{j\beta}\Gamma_4=
\frac{1}{4}\sum_{ijkl}\Biggl[K_{\alpha\beta}({\bf r}_i-{\bf r}_j)
\nonumber \\&& -K_{\alpha\beta}({\bf r}_k-{\bf
r}_j)\Biggr]\nabla_{i\alpha}\nabla_{j\beta}F_2({\bf r}_i-{\bf
r}_k)F_2({\bf r}_j-{\bf r}_l),
\end{eqnarray}
where the sum over $ijkl$ signifies the sum over all $24$
permutations of $1234$. The boundary conditions on $\Gamma_4$ are
vanishing with the difference of any of its arguments $|{\bf
r}_i-{\bf r}_j|$ tending to infinity. Equations on the irreducible
correlation functions of order higher than $4$ involve lower order
correlation functions. Let us consider the equation on $\Gamma_6$
defined by
\begin{eqnarray}&&
F_6=F^G_6+\sum_{i>j}F_2({\bf r}_i-{\bf r}_j)\Gamma_4(\{ {\bf
r}_{k\neq i, j} \})+\Gamma_{6}.
\end{eqnarray}
Using the equation on $\Gamma_4$ one finds for $\Gamma_6$
\begin{eqnarray}&&
{\cal L}_6 \Gamma_6=\frac{1}{3}\sum_{i>j, k\neq i, j}
\left[K_{\alpha\beta}({\bf r}_i-{\bf r}_k)-K_{\alpha\beta}({\bf
r}_j-{\bf r}_k \right]\nabla_{i\alpha} \nonumber
\\&& \nabla_{k\beta}F_2({\bf
r}_{i}-{\bf r}_{j})F^G_{4}(\{ {\bf r}_{k\neq i, j} \}) +\sum_{i>j,
k\neq i, j} \Biggl[K_{\alpha\beta}({\bf r}_i-{\bf r}_k)\nonumber
\\&&-K_{\alpha\beta}({\bf r}_j-{\bf r}_k
\Biggr]\nabla_{i\alpha}\nabla_{k\beta}F_2({\bf r}_{i}-{\bf
r}_{j})\Gamma_{4}(\{ {\bf r}_{k\neq i, j} \}).
\end{eqnarray}
One observes that the term that involves $\Gamma_4$ is of the
second order in non-Gaussianity that is at small $\xi$ it vanishes
as $\xi^2$. Thus this term is negligible to linear order in $\xi$
and in this order pair-correlation function fully determines the
source in the equation. This holds for $\Gamma_{n}$ with $n>6$ as
well. Indeed, to linear order in $\xi$ one can neglect all
products of $\Gamma-$s that enter the definition of $\Gamma_{n}$
(say, in the definition of $\Gamma_8$ one can neglect
$\Gamma_4\times \Gamma_4$ term. To linear order in $\xi$ the
definition of $\Gamma_{n}$ reads
\begin{eqnarray}&&
F_{n}=F_{n}^G+\Gamma_{n}+\sum_{k=2}^{n/2-1}\sum_{\{i_j\}}
\frac{\Gamma_{2k}\left(\{{\bf r}_{i_{1, 2, ..,2k}}
\}\right)}{(2k)!(n-2k)!}\nonumber \\&& \times
F^G_{n-2k}\left(\{{\bf r}_{i_{2k+1, ..,n}}\}\right),
\end{eqnarray}
where $\sum_{\{i_j\}}$ runs over all permutations of $1, 2,..,2n$
and the factor $[(2k)!(2n-2k)!]^{-1}$ lifts multiple counting of
terms. Using the equation on $F_{2n}$ it is easy to see that the
equation on $\Gamma_{2n}$ to linear order in $\xi$ coincides with
\begin{eqnarray}&&
{\cal L}_{2n}\Gamma_{2n}+\frac{n-1}{n}\sum_{i>j, k\neq i, j}
\nabla_{i\alpha}\nabla_{k\beta}\Biggl[K_{\alpha\beta}({\bf
r}_i-{\bf r}_k) \nonumber \\&& -K_{\alpha\beta}({\bf r}_j-{\bf
r}_k \Biggr]F_2({\bf r}_{i}-{\bf r}_{j})F^G_{2n-2}(\{ {\bf
r}_{k\neq i, j} \}) \nonumber \\&&
+\sum_{k=2}^{n-1}\sum_{\{i_j\}}\frac{1}{(2k)!(2n-2k)!}{\cal
L}_{2k}\Gamma_{2k}\left(\{{\bf r}_{i_{1, 2, ..,2k}} \}\right)
\nonumber \\&& F^G_{2n-2k}\left(\{{\bf r}_{i_{2k+1,
..,2n}}\}\right)=0.
\end{eqnarray}
Thus the equation satisfied by $\Gamma_{2n}$ can be written as
follows
\begin{eqnarray}&&
{\cal
L}_{2n}\Gamma_{2n}+C(n)\sum_{\{i_j\}}\nabla_{i_1\alpha}\nabla_{i_3\beta}
\Biggl[K_{\alpha\beta}({\bf r}_{i_1}-{\bf r}_{i_3}) \nonumber \\&&
-K_{\alpha\beta}({\bf r}_{i_2}-{\bf r}_{i_3}) \Biggr]\prod_{k=1}^n
F_2\left({\bf r}_{i_{2k-1}}-{\bf r}_{i_{2k}}\right)=0.
\nonumber\end{eqnarray} The value of the constant $C(n)$ is not
essential for our purposes. To derive the equation in the first
order in $\xi$ term $\Gamma_{2n}^1\equiv \lim_{\xi\to
0}\Gamma_{2n}/\xi$ we substitute ${\cal L}_{2n}$ by its value at
$\xi=0$. Using the fact that
$\sum_{i>j}\nabla_{i\alpha}\nabla_{j\alpha}=-\sum \nabla_i^2/2$
when acting on translation-invariant functions we find
\begin{eqnarray}&&
\sum_n\nabla_n^2\Gamma_{2n}^1 =F(n)
\sum_{\{i_j\}}\nabla_{i_1\alpha}\nabla_{i_3\beta}\Biggl[J_{\alpha\beta}
({\bf r}_{i_1}-{\bf r}_{i_3}) \nonumber \\&& +J_{\alpha\beta}({\bf
r}_{i_2}-{\bf r}_{i_4})-J_{\alpha\beta}({\bf r}_{i_2}-{\bf
r}_{i_3})-J_{\alpha\beta}({\bf r}_{i_1}-{\bf r}_{i_4}) \Biggr]
\nonumber \\&& \times\prod_{k=1}^n H\left({\bf r}_{i_{2k-1}}-{\bf
r}_{i_{2k}}\right), \label{cum1}\end{eqnarray} where $F(n)$ is an
$n$-dependent constant, $H=F_2(\xi=0)$ and
$J_{\alpha\beta}=\partial_{\xi}K_{\alpha\beta}({\bf r},
\xi)|_{\xi=0}$. Since constant part of $J_{\alpha\beta}$ cancels
in (\ref{cum1}) we will use the equivalent expression
\begin{eqnarray}&&
J_{\alpha\beta}({\bf r})=D(d-1)\ln r\delta_{\alpha\beta}-D
\frac{r_{\alpha}r_{\beta}}{r^2}.
\end{eqnarray}

\subsection{Solution of the equation on $\Gamma_{2n}^1$ and its large-scale asymptotic
expansion}

The solution can be expressed with the help of an auxiliary tensor
$T_{\alpha\beta}$ defined by
\begin{eqnarray}&&
\sum_{n=1}^{2n} \nabla^2_n T_{\alpha\beta}({\bf r}_1, {\bf
r}_2,..,{\bf r}_{2n}, {\bf x}_1,..,{\bf x}_k)=J_{\alpha\beta}({\bf
r}_{1}-{\bf r}_{3}) \nonumber \\&& \times
 \prod_{k=1}^n\delta\left({\bf r}_{2k-1}-{\bf
r}_{2k}-{\bf x}_k\right).
\end{eqnarray}
One has
\begin{eqnarray}&&
\Gamma_{2n}^1=F(n)\sum_{\{i_j\}}\nabla_{i_1\alpha}\nabla_{i_3\beta}\int
\prod_{k=1}^n H({\bf x}_k)\prod_{k=1}^n d{\bf x}_k \nonumber \\&&
\Biggl[T_{\alpha\beta}({\bf r}_{i_1}, {\bf r}_{i_2}, {\bf
r}_{i_3}, {\bf r}_{i_4}, {\bf r}_{i_5}..,{\bf r}_{i_{2n}}, {\bf
x}_1, {\bf x}_2, {\bf x}_3,..,{\bf x}_k)  \nonumber
\\&& +T_{\alpha\beta}({\bf r}_{i_2}, {\bf r}_{i_1}, {\bf r}_{i_4},
{\bf r}_{i_3}, {\bf r}_{i_5}..,{\bf r}_{i_{2n}}, -{\bf x}_1, -{\bf
x}_2, {\bf x}_3,..,{\bf x}_k)\nonumber
\\&& -T_{\alpha\beta}\Biggl({\bf r}_{i_2}, {\bf r}_{i_1}, {\bf
r}_{i_3}, {\bf r}_{i_4}, {\bf r}_{i_5}..,{\bf r}_{i_{2n}}, -{\bf
x}_1, {\bf x}_2, {\bf x}_3,..,{\bf x}_k\Biggr)\nonumber \\&&
-T_{\alpha\beta}({\bf r}_{i_1}, {\bf r}_{i_2}, {\bf r}_{i_4}, {\bf
r}_{i_3}, {\bf r}_{i_5}..,{\bf r}_{i_{2n}}, {\bf x}_1, -{\bf x}_2,
{\bf x}_3,..,{\bf x}_k) \Biggr]. \nonumber\end{eqnarray} Using
\begin{eqnarray}&&
\nabla^2 \frac{\Gamma(d/2-1)}{4\pi^{d/2} |{\bf r}-{\bf
r}'|^{d-2}}=-\delta\left({\bf r}-{\bf r}'\right)
\end{eqnarray}
we find
\begin{eqnarray}&&
-T_{\alpha\beta}=\int\frac{J_{\alpha\beta}({\bf r}'_{1}-{\bf
r}'_{3}) \prod_{k=1}^n\delta\left({\bf r}'_{2k-1}-{\bf
r}'_{2k}-{\bf x}_k\right)}{\left[\sum_1^{2n}({\bf r}_i-{\bf
r}'_i)^2\right]^{nd-1}}\nonumber \\&& \times
\frac{(nd-2)!}{4\pi^{nd}}\prod_{i=1}^{2n} d{\bf r}'_i=
\frac{\Gamma((n+2)d/2-1)}{2^{2+(n-2)d/2}\pi^{(n+2)d/2}} \int
\prod_{i=1}^{4} d{\bf r}'_i \nonumber
\\&& \frac{ J_{\alpha\beta}({\bf
r}'_{1}-{\bf r}'_{3}) \delta\left({\bf r}'_{1}-{\bf r}'_{2}-{\bf
x}_1\right)\delta\left({\bf r}'_{3}-{\bf r}'_{4}-{\bf
x}_2\right)}{\left[\sum_{i=1}^4({\bf r}_i-{\bf
r}'_i)^2+\sum_{k=3}^n\left({\bf r}_{2k-1}-{\bf r}_{2k}-{\bf
x}_k\right)^2\right]^{(n+2)d/2-1}}, \nonumber \end{eqnarray} where
we used
\begin{eqnarray}&&
\frac{\Gamma(n/2-1)}{4\pi^{n/2}}\int\frac{\prod_{i=1}^m dx'_i}
{\left[\sum_{i=1}^n(x_i-x_i')^2\right]^{n/2-1}} \nonumber \\&&
=\frac{\Gamma((n-m)/2-1)} {4\pi^{(n-m)/2}\left[\sum_{i=m+1}^n
(x_i-x_i')^2\right]^{(n-m)/2-1}}. \nonumber\end{eqnarray} which is
readily proved as a relation between Green's functions of a
laplacian in different dimensions. Introducing auxiliary vectors
${\bf z}_1={\bf r}_1,\ \ {\bf z}_2={\bf r}_2+{\bf x}_1,\ \ {\bf
z}_3={\bf r}_3-{\bf r},\ \ {\bf z}_4={\bf r}_4+{\bf x}_2-{\bf r}$,
where ${\bf r}={\bf r}'_1-{\bf r}'_3$ we have
\begin{eqnarray}&&
-T_{\alpha\beta}=\frac{\Gamma((n+2)d/2-1)}{2^{2+(n-2)d/2}\pi^{(n+2)d/2}}
\int d{\bf r}J_{\alpha\beta}({\bf r})\int d{\bf r}'_1\nonumber
\\&& \Biggl[\left(2{\bf r}'_1-\sum {\bf
z}_i/2\right)^2+\sum z_i^2-\left(\sum {\bf z}_i\right)^2/4
\nonumber
\\&&
+\sum_{k=3}^n\left({\bf r}_{2k-1}-{\bf r}_{2k}-{\bf x}_k\right)^2
\Biggr]^{1-(n+2)d/2}.\end{eqnarray} The integral over ${\bf r}'_1$
is easily calculated and we find
\begin{eqnarray}&&
-T_{\alpha\beta}=\frac{\Gamma((n+1)d/2-1)}{2^{3-d/2}\pi^{(n+1)d/2}}\int
d{\bf r}J_{\alpha\beta}({\bf r}) \nonumber
\\&& \times
\left[\left({\bf r}-{\bf b}\right)^2+c^2\right]^{1-(n+1)d/2},
\end{eqnarray}
where we introduced ${\bf b}={\bf r}_3+{\bf r}_4+{\bf x}_2-{\bf
r}_1-{\bf r}_2-{\bf x}_1$ and $c^2=\sum_{k=1}^n\left({\bf
r}_{2k-1}-{\bf r}_{2k}-{\bf x}_k\right)^2$. It is convenient to
introduce
\begin{eqnarray}&&
F_{\alpha\beta}({\bf b}, c^2, a)=\Gamma(a)\int d{\bf
r}\frac{J_{\alpha\beta}({\bf r})}{\left[\left({\bf r}-{\bf
b}\right)^2+c^2\right]^a} \nonumber
\\&&
-\Gamma(a)\int d{\bf r}\frac{J_{\alpha\beta}({\bf
r})}{\left[r^2+c^2\right]^a} \label{B7}\end{eqnarray} In terms of
this tensor the answer takes the form
\begin{eqnarray}&&
\Gamma_{2n}^1=\sum_{\{i_j\}}\frac{F(n)\nabla_{i_1\alpha}
\nabla_{i_3\beta}}{2^{3-d/2}\pi^{(n+1)d/2}} \int \prod_{k=1}^n
H({\bf x}_k) d{\bf x}_k \nonumber
\\&& \times\Biggl[F_{\alpha\beta}\Biggl(\frac{{\bf z}_{i_1, i_2,
i_3, i_4}+{\bf x}_1+{\bf x}_2}{2}, c^2_{\{i_j\}},
\frac{(n+1)d}{2}-1\Biggr) \nonumber \\&& +
F_{\alpha\beta}\Biggl(\frac{{\bf z}_{i_1, i_2, i_3, i_4} -{\bf
x}_1-{\bf x}_2}{2}, c^2_{\{i_j\}}, \frac{(n+1)d}{2}-1\Biggr)
\nonumber \\&& -F_{\alpha\beta}\Biggl(\frac{{\bf z}_{i_1, i_2,
i_3, i_4}+{\bf x}_1-{\bf x}_2}{2}, c^2_{\{i_j\}},
\frac{(n+1)d}{2}-1\Biggr) \nonumber \\&&
-F_{\alpha\beta}\Biggl(\frac{{\bf z}_{i_1, i_2, i_3, i_4}-{\bf
x}_1+{\bf x}_2}{2}, c^2_{\{i_j\}}, \frac{(n+1)d}{2}-1\Biggr)
\Biggr], \nonumber\end{eqnarray} where we introduced ${\bf
z}_{i_1, i_2, i_3, i_4}\equiv {\bf r}_{i_1}+{\bf r}_{i_2}-{\bf
r}_{i_3}-{\bf r}_{i_4}$ and $2c^2_{\{i_j\}}\equiv
\sum_{k=1}^n\left({\bf r}_{i_{2k-1}}-{\bf r}_{i_{2k}}-{\bf
x}_k\right)^2$. Using
\begin{eqnarray}&&
\frac{\partial F_{\alpha\beta}({\bf b}, c^2, a)}{\partial
c^2}=-F_{\alpha\beta}({\bf b}, c^2, a+1),\ \ \frac{\partial
F_{\alpha\beta}}{\partial b_{\beta}}=0 \nonumber\end{eqnarray} we
may rewrite the answer as
\begin{eqnarray}&&
\Gamma_{2n}^1=\frac{F(n)}{2^{2+nd/2}\pi^{(n+1)d/2}}\sum_{\{i_j\}}\int
\prod_{k=1}^n H({\bf x}_k)d{\bf x}_k \nonumber \\&& \times
\left({\bf r}_{i_{1}}-{\bf r}_{i_{2}}-{\bf
x}_1\right)_{\alpha}\left({\bf r}_{i_{3}}-{\bf r}_{i_{4}}-{\bf
x}_2\right)_{\beta} \nonumber
\\&& \times\Biggl[F_{\alpha\beta}\Biggl(\frac{{\bf z}_{i_1, i_2,
i_3, i_4}+{\bf x}_1+{\bf x}_2}{2}, c^2_{\{i_j\}},
\frac{(n+1)d}{2}+1\Biggr) \nonumber \\&& +
F_{\alpha\beta}\Biggl(\frac{{\bf z}_{i_1, i_2, i_3, i_4} -{\bf
x}_1-{\bf x}_2}{2}, c^2_{\{i_j\}}, \frac{(n+1)d}{2}+1\Biggr)
\nonumber \\&& -F_{\alpha\beta}\Biggl(\frac{{\bf z}_{i_1, i_2,
i_3, i_4}+{\bf x}_1-{\bf x}_2}{2}, c^2_{\{i_j\}},
\frac{(n+1)d}{2}+1\Biggr) \nonumber \\&&
-F_{\alpha\beta}\Biggl(\frac{{\bf z}_{i_1, i_2, i_3, i_4}-{\bf
x}_1+{\bf x}_2}{2}, c^2_{\{i_j\}}, \frac{(n+1)d}{2}+1\Biggr)
\Biggr] . \nonumber\end{eqnarray} Let us now derive the asymptotic
form of $\Gamma_{2n}^1$ at large distances for the case $q_0=0$.
In this case $H({\bf x})$ decays fast at $|{\bf x}|>L$ and the
expansion is derived in exactly the same way as the multi-pole
expansion of the potential in electrostatics. We find
\begin{eqnarray}&&
\Gamma_{2n}^1\approx \frac{F(n)\left(\int H({\bf x})x^2d{\bf
x}\right)^2\left(\int H({\bf x})d{\bf
x}\right)^{n-2}}{2^{2+nd/2}d^2\pi^{(n+1)d/2}} \nonumber \\&&
\sum_{\{i_j\}} \left({\bf r}_{i_{1}}-{\bf
r}_{i_{2}}\right)_{\alpha} \left({\bf r}_{i_{1}}-{\bf
r}_{i_{2}}\right)_{\gamma}\left({\bf r}_{i_{3}}-{\bf
r}_{i_{4}}\right)_{\beta}\left({\bf r}_{i_{3}}-{\bf
r}_{i_{4}}\right)_{\delta} \nonumber \\&& \times \frac{\partial^2
F_{\alpha\beta}}{\partial b_{\gamma}\partial
b_{\delta}}\Biggl(\frac{{\bf z}_{i_1, i_2, i_3, i_4}}{2},
c^2_{\{i_j\}}, \frac{(n+1)d}{2}+3\Biggr).
 \nonumber \end{eqnarray}
note that $$\int H({\bf x})x^{k} d{\bf x}=-[(k+2)(k+d)D]^{-1} \int
\chi({\bf x})x^{k+2}d{\bf x}\ .$$ To complete the calculation we
pass to study $F_{\alpha\beta}$.

\subsection{Computation of $F_{\alpha\beta}$}

We now concentrate on $F_{\alpha\beta}$ defined by Eq. (\ref{B7}).
By isotropy and incompressibility $F_{\alpha\beta}$ satisfies
\begin{eqnarray}&&
F_{\alpha\beta}({\bf b})=A(b)\delta_{\alpha\beta}+B(b){\hat
b}_{\alpha}{\hat b}_{\beta},\nonumber  \\&& \frac{\partial
F_{\alpha\beta}}{\partial b_{\beta}}=0=A'+B'+\frac{B(d-1)}{b}.
\end{eqnarray}
We consider the part of $F_{\alpha\beta}$ that enters
$\Gamma_{2n}^1$. Differentiating the above equation we find that
for any two vectors ${\bf v}$, ${\bf w}$ we have
\begin{eqnarray}&&
v_{\alpha}v_{\gamma}w_{\beta}w_{\delta}\frac{\partial^2
F_{\alpha\beta}({\bf b}, c^2, a)}{\partial b_{\gamma}\partial
b_{\delta}}=\frac{Bv^2w^2}{b^2}- \left({\bf v}\cdot {\bf
w}\right)^2 \nonumber \\&& \times
\left(\frac{B'}{b}+\frac{(d-2)B}{b^2}\right)+\left(\frac{B'}{b^3}-\frac{2B}{b^4}\right)
\Biggl[v^2\left({\bf w}\cdot {\bf b}\right)^2\nonumber
\\&&
+w^2\left({\bf v}\cdot {\bf b}\right)^2\Biggr]
+\left(b\left(\frac{B'}{b}\right)'-4\frac{B'}{b}
+\frac{8B}{b^2}\right)\frac{\left({\bf v}\cdot {\bf
b}\right)^2}{b^4} \nonumber
\\&& \times \left({\bf w}\cdot {\bf b}\right)^2
-\left(b\left(\frac{B'}{b}\right)'+\frac{(d-4)B'}{b}-\frac{2(d-4)
B^2}{b^2}\right) \nonumber
\\&& \times\frac{\left({\bf v}\cdot {\bf w}\right)\left({\bf
v}\cdot {\bf b}\right)\left({\bf w}\cdot {\bf b}\right)}{b^2}
\end{eqnarray}
Thus our purpose is to find $B$. Taking the trace we find the
additional equation that connects $A$ and $B$:
\begin{eqnarray}&&
dA+B=Dd\Gamma(a) \left[I(b)-I(b=0)\right],\nonumber \\&&
I(b)\equiv\int d{\bf r}\frac{\ln r}{\left[\left({\bf r}-{\bf
b}\right)^2+c^2\right]^{a}}. \nonumber\end{eqnarray} To perform
angular integration conveniently we note that $I({\bf b})$ is a
convolution of two spherically symmetric functions and its Fourier
image is a product of spherically symmetric functions. To use
Fourier transform techniques with logarithmic function we write
$\ln r=-\partial r^{-\xi}/\partial \xi(\xi=0)$. Using
\begin{eqnarray}&&
\int d{\bf b}\frac{\exp\left[i{\bf b}\cdot{\bf
q}\right]}{\left(b^2+c^2\right)^a}=\frac{(2\pi)^{d/2}
\left(cq\right)^{d/2-a}
K_{d/2-a}(cq)}{2^{a-1}\Gamma(a)q^{d-2a}},\nonumber \\&& \ \int
d{\bf b} e^{i{\bf b}\cdot{\bf
q}}b^{-\xi}=\frac{(2\pi)^{d/2}2^{d/2-\xi}\Gamma((d-\xi)/2)}
{q^{d-\xi}\Gamma(\xi/2)}, \nonumber\end{eqnarray} where $K_{\nu}$
is a modified Bessel function of order $\nu=a-d/2$, we find
\begin{eqnarray}&&
I(b)=\left(-\frac{\partial}{\partial
\xi}\right)|_{\xi=0}\frac{(2\pi)^{d/2}c^{-\nu}\Gamma((d-\xi)/2)}{b^{d/2-1}
2^{\nu+\xi-1}\Gamma(\xi/2)\Gamma(a)}\nonumber
\\&&
\int_0^{\infty} q^{a+\xi-d}J_{d/2-1}(qb)K_{\nu}(cq)dq
=\left(-\frac{\partial}{\partial \xi}\right)|_{\xi=0}\nonumber
\\&&\frac{\Gamma(\nu+\xi/2)\Gamma((d-\xi)/2)\pi^{d/2}}
{c^{2\nu+\xi}\Gamma(d/2)\Gamma(a)}F\left(\nu+\frac{\xi}{2},
\frac{\xi}{2}, \frac{d}{2}, -\frac{b^2}{c^2}\right),
\nonumber\end{eqnarray} where $J_{d/2-1}$ is Bessel function of
order $d/2-1$ and $F$ is a hypergeometric function. Using $I(b)$
and $I(0)$ we find
\begin{eqnarray}&&
dA+B=\frac{D d
\Gamma(\nu)\pi^{d/2}}{2c^{2\nu}}\left(-\frac{\partial}{\partial
\xi}\right)|_{\xi=0} \nonumber \\&& F\left(\nu, \xi, \frac{d}{2},
-\frac{b^2}{c^2}\right), \nonumber\end{eqnarray} where we used
properties of hypergeometric function. One easily finds that $B$
satisfies
\begin{eqnarray}&&
\frac{1}{b^d}\frac{\partial b^dB}{\partial b}=\frac{Dd
\Gamma(\nu)\pi^{d/2}}{2c^{2\nu}}\left(\frac{\partial}{\partial
b}\right)\left(\frac{\partial}{\partial \xi}\right)|_{\xi=0}
\nonumber \\&& F\left(\nu, \xi, \frac{d}{2},
-\frac{b^2}{c^2}\right).
\end{eqnarray}
The result of the differentiation over $\xi$ can be written as a
series using
$$\left(\frac{\partial (\xi)_k}{\partial
\xi}\right)_{\xi=0}=(k-1)!\ ,$$ so that
\begin{eqnarray}&&\left(\frac{\partial}{\partial
\xi}\right)|_{\xi=0}F\left( \nu, \xi, \frac{d}{2},z\right)=
\sum_{k=1}^\infty\frac{(\nu)_kz^k}{k(d/2)_k}\ .
\end{eqnarray}
We find for $B$
\begin{eqnarray}&&
B=\frac{Dd\Gamma(\nu)\pi^{d/2}}{2c^{2\nu}(d-1)}\sum_{k=1}^\infty
\frac{(\nu)_k}{(d/2)_k(d/2+k)} \left(-\frac{b^2}{c^2}\right)^{k}
\nonumber
\\&&
=-\frac{2D\Gamma(\nu+1)\pi^{d/2}b^2}{(d-1)(d+2)c^{2\nu+2}}
F\left(\nu+1, 1, \frac{d}{2}+2, -\frac{b^2}{c^2}\right).
\nonumber\end{eqnarray} Next we find
\begin{eqnarray}&&
\frac{B'}{b}\!=\!-\frac{4D\Gamma(\nu+1)\pi^{d/2}}{(d-1)(d+2)c^{2\nu+2}}
F\left(\nu+1, 2, \frac{d}{2}+2, -\frac{b^2}{c^2}\right).
\nonumber\end{eqnarray} It follows using
\begin{eqnarray}&&
\frac{\partial F(\alpha, \beta, \gamma, z)}{\partial
z}=\frac{\alpha\beta}{\gamma}F(\alpha+1, \beta+1, \gamma+1, z)
\nonumber\end{eqnarray} that
\begin{eqnarray}&&
b\left(\frac{B'}{b}\right)'=
\frac{32D\Gamma(\nu+2)\pi^{d/2}}{(d-1)(d+2)(d+4)c^{2\nu+2}}
\left(\frac{b^2}{c^2}\right)\nonumber \\&& \times F\left(\nu+2, 3,
\frac{d}{2}+3, -\frac{b^2}{c^2}\right).
\end{eqnarray}
Combining the results of the last two subsections one finds the
answer for $\Gamma_{2n}^2$ given in the main text.

\end{document}